\documentclass[aps,pra,longbibliography,twocolumn,showpacs,nofootinbib,superscriptaddress,notitlepage]{revtex4-2}
\usepackage{amsmath}
\usepackage{amssymb,bm}
\usepackage{amsthm}

\usepackage{qcircuit}

\newtheorem{innercustomgeneric}{\customgenericname}
\providecommand{\customgenericname}{}
\newcommand{\newcustomtheorem}[2]{%
  \newenvironment{#1}[1]
  {%
   \renewcommand\customgenericname{#2}%
   \renewcommand\theinnercustomgeneric{##1}%
   \innercustomgeneric
  }
  {\endinnercustomgeneric}
}

\newcustomtheorem{customthm}{Theorem}
\newcustomtheorem{definitionBob}{Definition}

\usepackage{color,dsfont} 
\usepackage{graphicx}
\usepackage{subcaption}
\usepackage{ragged2e}
\DeclareCaptionJustification{justified}{\justifying}
\usepackage[colorlinks=true, hyperindex, breaklinks, linkcolor=blue, urlcolor=blue, citecolor=blue]{hyperref} 
\usepackage[normalem]{ulem}
\usepackage[capitalise]{cleveref}

\usepackage{mathrsfs}

\usepackage{mathtools}
\usepackage{float}
\usepackage{verbatim}
\usepackage{latexsym}
\usepackage{amsmath}
\usepackage{amssymb}
\usepackage{setspace}
\usepackage{amsfonts}
\usepackage{stmaryrd}
\usepackage{xcolor}
\usepackage{enumitem}
\usepackage{hhline}
\usepackage[normalem]{ulem}

\crefname{lemma}{Lemma}{Lemmas}

\newcommand{\bra}[1]{\langle#1|} 
\newcommand{\ket}[1]{|#1\rangle} 
 
\newcommand{\codepar}[1]{\ensuremath{[\![#1]\!]}}

\crefname{equation}{Eq.\!}{Eqs.\!}
\crefname{figure}{Fig.\!}{Figs.\!}
\usepackage{multirow}
\mathchardef\mhyphen="2D

\begin{document}

\title{Logical coherence in 2D compass codes}

\author{Balint Pato}
\email{balint.pato@duke.edu}
\affiliation{
  Duke Quantum Center, Duke University, Durham, NC 27701, USA
}
\affiliation{
  Department of Electrical and Computer Engineering, Duke University, Durham, NC 27708, USA
}

\author{J. Wilson Staples}

\affiliation{
  Duke Quantum Center, Duke University, Durham, NC 27701, USA
}
\affiliation{
  Department of Physics, Duke University, Durham, NC 27708, USA
}
\affiliation{
  Department of Electrical and Computer Engineering, Duke University, Durham, NC 27708, USA
}

\author{Kenneth R. Brown}

\affiliation{
  Duke Quantum Center, Duke University, Durham, NC 27701, USA
}
\affiliation{
  Department of Electrical and Computer Engineering, Duke University, Durham, NC 27708, USA
}
\affiliation{
  Department of Physics, Duke University, Durham, NC 27708, USA
}
\affiliation{
  Department of Chemistry, Duke University, Durham, NC 27708, USA
}

\begin{abstract}
  2D compass codes are a family of quantum error-correcting codes that contain the Bacon-Shor codes, the $X$-Shor and $Z$-Shor codes, and the rotated surface codes. Previous numerical results suggest that the surface code has a constant accuracy and coherence threshold under uniform coherent rotation. However, having analytical proof supporting a constant threshold is still an open problem. It is analytically proven that the toric code can exponentially suppress logical coherence in the code distance $L$. However, the current analytical lower bound on the threshold for the rotation angle $\theta$ is $|\sin(\theta)| < 1/L$, which linearly vanishes in $L$ instead of being constant. We show that this lower bound is achievable by the $Z$-Shor code which does not have a threshold under stochastic noise. Compass codes provide a promising direction to improve on the previous bounds. We analytically determine thresholds for two new compass code families that provide upper and lower bounds to the rotated surface code's numerically established infidelity threshold. Furthermore, using a Majorana mode-based simulator, we use random families of compass codes to smoothly interpolate between the $Z$-Shor codes and the $X$-Shor codes.
\end{abstract}


\maketitle

\section{Introduction}

The main obstacle in unlocking the power of quantum computers is overcoming noise, a collection of unwanted physical processes in the system. Quantum error correction threshold theorems \cite{aharonov_fault_1996,DKLP02} state that under probabilistic (incoherent) models of quantum noise, the rate of error can be suppressed exponentially by encoding the quantum information into increasingly large quantum error-correcting codes and carefully curbing the propagation of faults \cite{Shor96}. Quantum error correction threshold theorems apply only to a specific code family construction and a specific noise model. The most celebrated analytical threshold results assume incoherent noise for concatenated \cite{aharonov_fault_1996, aliferis_accuracy_2007} and topological code families \cite{DKLP02}. With increasing stabilizer code size, coherent (unitary) noise has also been shown to be suppressed \cite{Huang_Doherty_2019,beale_quantum_2018}, but an actual constant threshold has not been established, only a linearly vanishing lower bound in code size \cite{iverson_coherence_2020}. Numerical threshold results for coherent noise are limited to small systems or surface codes that can be represented with Majorana modes. This is because small rotations need either state vector \cite{gutierrez_errors_2016-1,Barnes_Trout_Lucarelli_Clader_2017} or tensor network simulation \cite{darmawan_tensor-network_2017} or limited Majorana mode simulators \cite{bravyi_correcting_2018,Venn_Beri_2020,Marton_Asboth_2023,Pataki_Marton_2024}, in contrast with wide categories of incoherent noise that can be simulated using the Gottesman-Knill theorem \cite{Gottesman98b}. Coherent noise models describe a non-stochastic, unitary evolution on the physical qubits. This kind of unwanted evolution arises from the imperfections of quantum control. The effect of coherent errors in FTQEC systems can be mitigated with randomized compilation \cite{Jain_Iyer_Bartlett_Emerson_2023,Wallman_Emerson_2016} and other techniques \cite{debroy_stabilizer_2018,cai_mitigating_2020, hu_mitigating_2021}. However, when the underlying assumptions of these techniques are not perfectly satisfied, residual coherent errors will still have an impact that we seek to understand. Beyond modeling coherent errors, coherent physical channels also gained interest in logical gate design \cite{hu_climbing_2021,hu_designing_2021,Hu_Liang_Calderbank_2022,Akahoshi_Maruyama_Oshima_Sato_Fujii_2023}.

In this paper, we continue the exploration of the threshold of surface codes under the uniform coherent rotation channel, $(e^{-i\frac{\theta}{2}Z})^{\otimes n}$ \cite{bravyi_correcting_2018, iverson_coherence_2020, Venn_Beri_2020, Venn_Behrends_Beri_2023}, with a focus on the rotated surface code \cite{Bombin_Martin-Delgado_2007}. Analytically, the best asymptotic result only guarantees a threshold for a linearly disappearing range of $\sin(\theta) < 1/L$ in code size $L=\sqrt{n}$ by Iverson and Preskill \cite{iverson_coherence_2020} for the toric code \cite{BK98}. However,  Bravyi et al. \cite{bravyi_correcting_2018} conjectured a constant threshold of $\theta \leq \frac{\pi}{5}$ under coherent noise based on numerical evidence using minimum weight perfect matching (MWPM) decoding \cite{DKLP02}. Venn, Behrends, and Béri \cite{Venn_Behrends_Beri_2023} found a threshold of $(0.28 \pm 0.01)\pi$ for the rotated surface code by numerically evaluating phases of an equivalent Majorana metal, which can be interpreted as a maximum likelihood (ML) decoding threshold. These constant numerical threshold results indicate that the analytical lower bound should have plenty of room for improvement. Our first result underlines this conclusion as it shows how the $Z$-Shor code, which does not have a threshold for stochastic noise, also suppresses the coherence of the logical channel in this regime, using ML decoding. In our second contribution, we define repetition code-like families of 2D compass codes, the $Z$-stacked Shor codes, with closed form analytical expressions for their thresholds using ML decoding. Finally, our third result uses MWPM to decode randomly generated 2D compass codes that smoothly interpolate on average between the disappearing threshold of the $Z$-Shor code and the maximal $\pi/2$ threshold of the $X$-Shor code. The random codes are evaluated numerically using a Majorana mode simulator which is based on the work of Bravyi et al. \cite{bravyi_correcting_2018} and is generalized to compass codes.

The paper is structured as follows. In \cref{sec: background}, we establish notation and review the logical channel under uniform coherent rotation and review 2D compass codes.  In \cref{sec:coherence thresholds} we derive coherence thresholds for repetition code like compass code families and showcase our numerical results. Finally, we conclude our findings in \cref{sec: conclusions}.

\def\thalf{\frac{\theta}{2}}

\section{Background} \label{sec: background}

We simulate logical quantum memory instead of computation and assume perfect state preparation and syndrome extraction. We focus on the performance of encodings of a single logical qubit into a \codepar{n,1,d} error-correcting code under the channel that consists of coherent noise, $X$-syndrome measurement, and recovery. This physical channel acts on the physical state $\rho$ as

\begin{align}
  \mathcal{E}(\rho)= \sum_s Z(h_s) \Pi_s \mathcal{N}_\theta(\rho) \Pi_s Z(h_s),
\end{align}

\noindent where $\Pi_s$ is the projector to the subspace compatible with measuring syndrome $s$, the recovery operator $Z(h_s)$ is defined by applying Pauli Z operators on the support of the bit-string $h_s \in \mathbb{F}_2^n$, and $\mathcal{N}_\theta(\rho)$ is the uniform Z rotation by angle $\theta$ on all $n$ physical qubits, $\mathcal{N}_\theta(\rho)=U_Z \rho U_Z^\dagger, \text{ with }  U_Z=(e^{-i \thalf Z})^{\otimes n}$.

We are interested in how much the logical channel differs from the identity channel and how this difference grows by applying the logical channel repeatedly $m$ times. It has been previously shown that in the limit of large code sizes, the logical channel of stabilizer codes is close to a stochastic channel, thus well characterized by the logical infidelity \cite{beale_quantum_2018}, and is worthwhile optimizing decoding for. Note that this is in contrast with physical channels, where gate fidelity might not be predictive of scalability \cite{sanders_bounding_2015, wallman_estimating_2015}. Our distance metric is the average infidelity of the logical states after $m$ applications of the logical channel, which we simply refer to as the infidelity of the logical channel:

\begin{align}
  r_m=1-\int_\psi d\psi \bra{\psi}\mathcal{E}^m(\ket{\psi}\bra{\psi})\ket{\psi}.
\end{align}

We can interpret a single application of the logical channel as a probabilistic mixture of logical rotations for \codepar{n,1,d} CSS codes with $k=1$ and $d$ odd \cite{bravyi_correcting_2018, iverson_coherence_2020}. Following \cite{iverson_coherence_2020}, for our $Z$-rotation-only case, we parameterize the Pauli Transfer Matrix (PTM) of the logical channel:

$$\bar{N}=\left(
  \begin{array}{cccc}
      1 & 0          & 0          & 0 \\
      0 & 1-\epsilon & \delta     & 0 \\
      0 & -\delta    & 1-\epsilon & 0 \\
      0 & 0          & 0          & 1 \\
    \end{array}
  \right),$$

\noindent where $\epsilon = \sum_s p_s (1 - \cos(\theta_s))$ and $\delta = \sum_s p_s \sin(\theta_s)$ are the averages of $1-\cos(\theta_s)$ and $\sin(\theta_s)$ over the distribution of logical rotation angles $\theta_s$ for syndromes $s$ with corresponding probabilities $p_s$. For a Hilbert-space with dimension $d=2^n$ for $n$ qubits, the PTM channel representation follows from representing the density matrix as a vector $\rho=\sum_{j=0}^{d^2-1} \rho_j \sigma^j$ in the Pauli basis $\{\rho^j\}$. Then the channel becomes a matrix $\bar{N}_{ij} = \text{Tr}[\mathcal{N}(\sigma^{j}) \sigma^{i}]$.

The fact that we can define a logical angle $\theta_s$ for a given syndrome $s$ comes from the Pauli-expansion of $U_Z$, under the assumptions that there is only a single logical $\bar{Z}$ operator ($k=1$) and that all the codes in our discussion have even weight stabilizers and an odd distance, giving rise to an odd-weight logical $\bar{Z}$. For a derivation, see Lemma 4 in \cite{bravyi_correcting_2018}.

For a single application of the logical channel, $r_1=\frac{\epsilon}{3}$ \cite{iverson_coherence_2020}, thus $\epsilon$ is proportional to infidelity. A non-zero $\delta$ indicates rotation between logical $X$ and $Y$. For example, when a Pauli channel is completely incoherent, then its PTM is diagonal, with $\delta=0$. For a completely coherent, single qubit rotation $U=e^{-i\theta/2Z}$, the diagonal element is $1-\epsilon = \cos(\theta)$, and the off-diagonal element is $\delta =\sin(\theta)$. As we'll see in \cref{sec:coherence thresholds}, only $\epsilon$ contributes to $r_1$, but both of these terms contribute to $r_m$. To leading order, $\epsilon$ contributes linearly, while $\delta$ contributes a quadratic term in $m$.

Analytical approaches to calculate distance metrics for a logical channel exactly under coherent noise are possible in general for smaller systems \cite{greenbaum_modeling_2018} with tensor network-based methods extending the reach by exploiting symmetries in the codes \cite{cao_quantum_2023}. As the number of terms grows exponentially with the system, approximations are required for asymptotic analysis and establishing lower bounds \cite{greenbaum_modeling_2018, iverson_coherence_2020}. For codes with a simpler structure, like repetition codes, one can get exact results to a larger size by evaluating a number of terms growing sub-exponentially with system size \cite{iverson_coherence_2020}.


\subsection{Logical channel of the repetition code} \label{sec: rep codes}

Consider a chain of qubits of length $d_z$. We define repetition codes (or phase flip codes) by their stabilizer group as $\mathcal{S}= \langle X_iX_{i+1} \mid i \in [1, d_Z-1] \rangle$. Iverson and Preskill \cite{iverson_coherence_2020} calculated closed-form approximations for the logical channel parameters $\epsilon$ and $\delta$ as functions of $\theta$:

\begin{align}
  \epsilon_{l}(\theta) & \approx \sqrt{\frac{2}{\pi l}}\frac{\sin^{l+1}\theta}{\cos \theta} (1+\mathcal{O}(\frac{1}{l})) \label{eq: approx_eps} \\
  \delta_{l}(\theta)   & \approx \frac{\cos \theta}{\sin \theta}\epsilon_{l}(\theta) (1+\mathcal{O}(\frac{1}{l})) \label{eq: approx_delta},
\end{align}

\noindent where the approximation is based on Stirling's formula, and valid when $\sin(\frac{\theta}{2})^2 < 1/2$, thus, when $|\theta| < \frac{\pi}{2}$. We heavily rely on these formulas below to understand quantum codes that can be decoded as products of repetition codes.

\subsection{2D compass codes}\label{ssec: 2d compass}

\begin{figure}[!htbp]
  \centering
  \captionsetup[subfigure]{justification=centering}
  \begin{subfigure}[b]{0.2\textwidth}
    \centering
    \includegraphics[width=\textwidth]{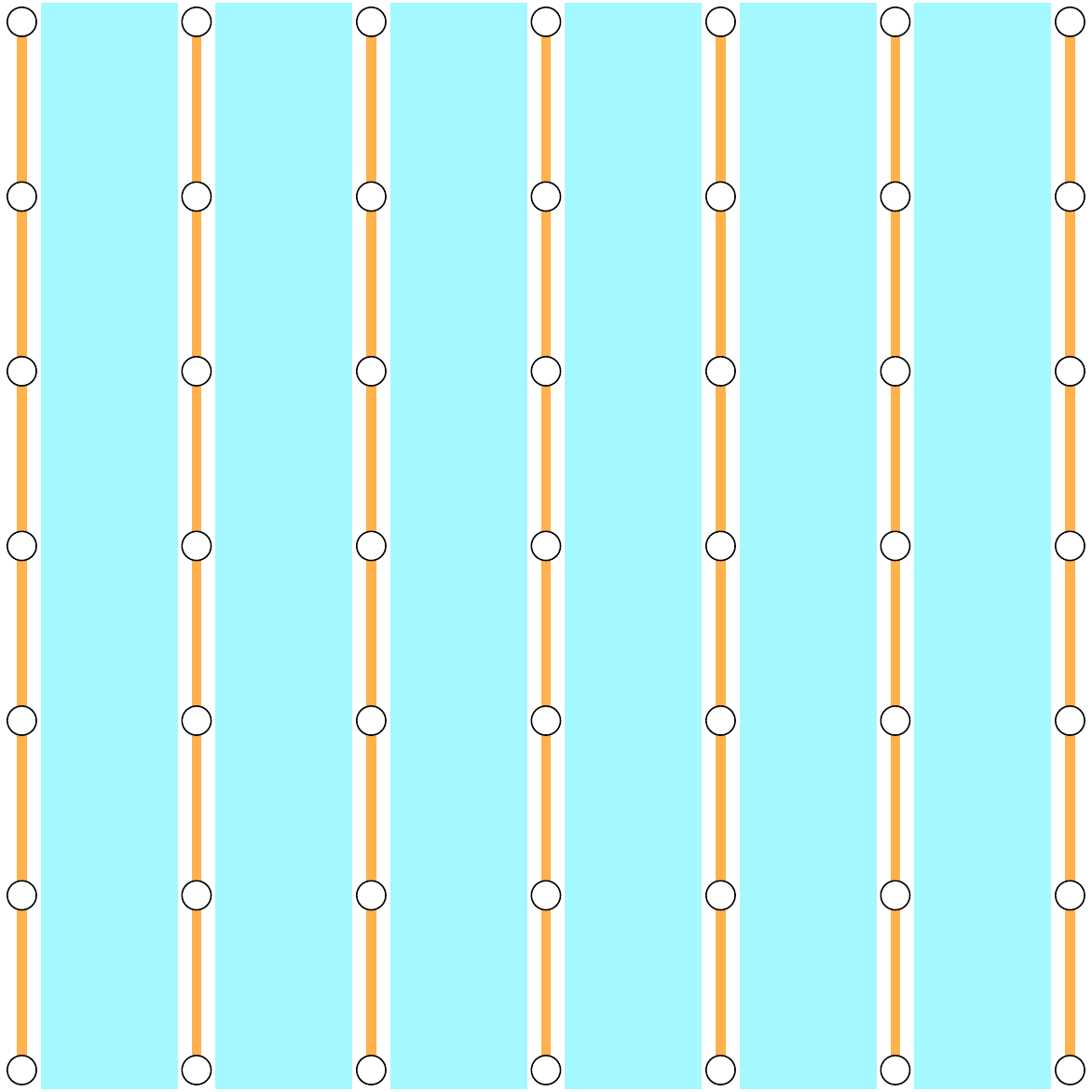}
    \caption{$Z$-Shor code}
    \label{fig:zshor}
  \end{subfigure}
  \begin{subfigure}[b]{0.2\textwidth}
    \centering
    \includegraphics[width=\textwidth]{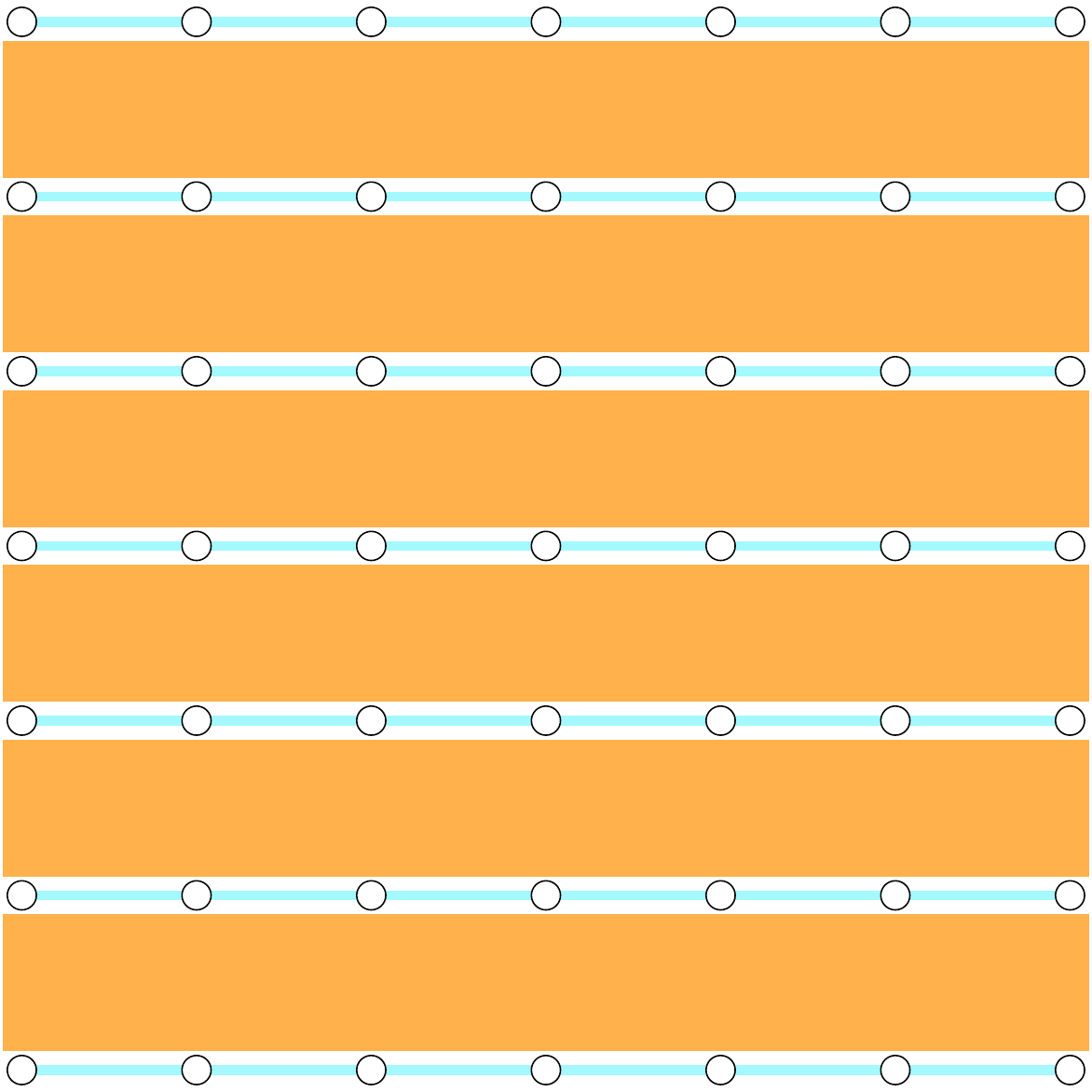}
    \caption{$X$-Shor code}
    \label{fig:xshor}
  \end{subfigure}
  \caption{ (color online) (a) The 5x5 Shor's code which we call $Z$-Shor code, referring to the type of the weight-2 stabilizer generators and (b) the 5x5 $X$-Shor code.}
  \label{fig: zxshor}
\end{figure}

2D compass codes \cite{Bacon06, li_2d_2019} are CSS codes \cite{CS96} closely related to the quantum compass model \cite{Kugel_Khomskii_1972,Dorier_Becca_Mila_2005}, that describes interacting spins on an $l\times l$ lattice described by a Hamiltonian

$$H=-\sum_{r } \sum_{c < l-1} J_X X_{r,c} X_{r,c+1} - \sum_{r < l-1} \sum_{c} J_Z Z_{r,c} Z_{r+1,c}, $$

\noindent where a Pauli operator $P_{r,c}$ acts on qubits in row $r$ and column $c$. Compass codes are derived from the Bacon-Shor subsystem code \cite{Bacon06, Shor95} based on $l^2$ physical qubits with a stabilizer group $S=\langle \prod_r X_{r,c} X_{r,c+1}, \prod_c Z_{r,c} Z_{r+1,c}\rangle$ of $2l-2$ stabilizer generators and one logical qubit, leaving $l^2-2l+1$ gauge degrees of freedom in the gauge group $\mathcal{G}=\langle X_{r,c} X_{r,c+1},Z_{r,c} Z_{r+1,c}\rangle$, which are generated by the two-body terms of the compass model. Gauge operators $g \in \mathcal{G}$ can be ``fixed'' \cite{Paetznick_Reichardt_2013, Bombin_2015} by moving $g$ to $S$ and removing all gauge operators from $\mathcal{G}$ that anti-commute with $g$. Compass codes are constructed through gauge fixing the Bacon-Shor code. In this paper, we also allow for rectangular compass codes; furthermore, we fix all gauge degrees of freedom, and thus our codes have $n-1$ stabilizer generators. We define compass codes on a $d_x \times d_z$ grid of qubits, $d_z$ odd (even distance codes have no coherence in their logical channel). Three known families of codes are the $Z$-Shor codes, $X$-Shor codes, and the rotated surface code \cite{Bombin_Martin-Delgado_2007}. We call Shor's code \cite{Shor96} the $Z$-Shor code, to emphasize the fact that the weight-2 stabilizer generators are $Z$-type, displayed in \cref{fig:zshor}. The Hadamard transform of the $Z$-Shor code is the $X$-Shor code, where the weight-2 stabilizers are $X$-type, displayed in \cref{fig:xshor}. It is useful to think of 2D compass codes as red ($Z$-type) and blue ($X$-type) colorings of a $(d_X-1) \times (d_Z-1)$ size grid \cite{li_2d_2019}, in which case, $Z$-Shor is all blue, and $X$-Shor is all red and the rotated surface code follows a checkerboard pattern.

\section{Analytical and numerical coherence thresholds}\label{sec:coherence thresholds}

\subsection{Coherence threshold for the repetition code}\label{ssec: rep code coh threshold}

We first introduce a new measure for coherence. To leading order, the growth of $r_m$ in $m$ is linear for stochastic channels but quadratic for purely coherent channels, as shown by Iverson and Preskill \cite{iverson_coherence_2020}. For the repetition code's logical channel, Iverson and Preskill derived the approximate expression for $r_m$ (when $\frac{m^2 \epsilon}{3}$ is small), which shows that $\delta$ drives the quadratic term:

\begin{align}
  r_m & = \frac{1}{3} \epsilon m - \frac{1}{6} m(m-1) \delta^2                          \\
      & +  \mathcal{O}(m^2\epsilon^2, m^3\epsilon^3, m^3 \epsilon\delta^2, m^4\delta^4)
\end{align}

It is now clear why $\epsilon$ is a measure of infidelity, as for a single repetition, $r_1 \approx \frac{1}{3} \epsilon$. Thus, we choose $\kappa(\theta) \equiv \frac{\delta(\theta)^2}{\epsilon(\theta)}$ as a simple measure for coherence representing the strength of the quadratic term in $r_m$ relative to the linear term. Then, the coherence of the logical channel of repetition codes, after plugging in the approximate $\epsilon$ and $\delta$ values described by \cref{eq: approx_eps,eq: approx_delta} is as follows:

\begin{align}
  \kappa_{l}(\theta) & \approx  \left(\frac{\cos \theta }{\sin \theta}\right)^2 \epsilon_{l}(\theta) (1+\mathcal{O}(\frac{1}{l}))
\end{align}

We define the coherence threshold $\theta_{th}$ to be the rotation value under which it is guaranteed that $\kappa_l(\theta)$ is suppressed exponentially in the code size $l$. The infidelity threshold $\theta_{th}^{r_1}$ is the rotation value for which all $\theta \leq \theta_{th}^{r_1}$, $r_1(\theta)$ is exponentially suppressed in the code size $l$.
For any $\theta < \frac{\pi}{2}$, both $\kappa_l(\theta)$ and $\epsilon_l(\theta)$  decrease exponentially in $l$, thus we conclude that the coherence threshold and the infidelity threshold are both $\theta_{th}^{rep} = \frac{\pi}{2}$ for the repetition code.

As minimum-weight decoding is equivalent to ML decoding for repetition codes under code capacity noise models, these repetition code thresholds can be considered to be both under ML and minimum-weight decoding. See \cref{app:verifying formulae} for further numerical evidence on this.

\subsection{Coherence threshold for the $Z$-Shor code} \label{ssec: $Z$-Shor code threshold}

The $d_x=1$ $Z$-Shor code is the length $d_z$ phase-flip code. For $d_x>1$, the fact that any logical state $\ket{\psi_L}$ is stabilized $Z_{i,j}Z_{i+1,j}$ causes the property that $e^{-i\thalf Z_{i,j}} \ket{\psi_L} = e^{-i\thalf Z_{i+1,j}} \ket{\psi_L} $. Using this, in every column $j$, every single qubit $Z$-rotation on row $i$ can be equivalent to acting on qubit $1,j$, resulting in an effective rotation of $e^{-i d_x \frac{\theta}{2} Z_{1,j}}$, which we refer to as ``zipping'' of the rotation along ZZ stabilizers.

Thus, the $Z$-Shor code under the uniform rotation $U_Z(\theta)$ has the same logical channel as the length $d_z$ phase-flip code under the uniform rotation $U_Z(d_x \theta)$. It follows that the logical channel parameters can be simply expressed as the phase-flip parameters at physical angle $d_x \theta$, resulting in the measure of coherence of the $Z$-Shor code being $\kappa_{Z, d_x \times d_z}(\theta) =  \kappa_{d_z}(d_x \theta)$. From here it is easy to see that for a given $d_x$ both $\kappa_{Z, d_x \times d_z}(\theta)$ and $\epsilon_{Z, d_x \times d_z}(\theta)$ are suppressed exponentially in $d_z$ when $\theta < \frac{\pi}{2 d_x}$. Interestingly, when $d_x = d_z = l$, for small $\theta$, this means that $\theta \approx \sin(\theta) < \frac{\pi}{2} 1/l$, the lower bound Iverson and Preskill derived for the toric code \cite{iverson_coherence_2020}. The $Z$-Shor code has maximal coherence, which we expect to be strictly worse than any compass code, including the surface code. Thus, our result points towards a need for a better lower bound for the surface code threshold.

Physical rotations above $\theta=\frac{\pi}{2 d_x}$ for the $Z$-Shor code are equivalent to being above the repetition code threshold of $\pi/2$, and thus the the minimum weight recovery becomes incorrect. An ML decoder \cite{Pato_Miao_etal_2024} would then find the correct recovery operator. Thus, in the case of the $Z$-Shor code, ML and minimum weight recovery are not the same and our repetition code-based threshold for the $Z$-Shor code should be interpreted to be under minimum-weight recovery. In \cref{app:verifying formulae} we provide further numerical evidence for this argument.

\subsection{Coherence threshold for the $X$-Shor code}

We can think of the $X$-Shor code as $d_x$ length-$d_z$ phase-flip codes glued together by weight-$2d_z$ Z operators. As such, the logical channel of the $X$-Shor code is equivalent to applying $d_x$ times the logical channel of the length-$d_z$ phase-flip code. In the Pauli Transfer Matrix representation, the combined PTM of applying multiple channels is simply the product of the individual channel PTMs. In our case, the PTM for the $X$-Shor code's logical channel is going to be the product of the PTMs of the phase-flip code logical channel.

To calculate the parameters, we diagonalize the PTM of the phase-flip code. The eigenvalues are 1, 1, and $1-\epsilon_{l}(\theta) \pm i\delta_{l}(\theta) \equiv \exp(\lambda_{l}(\theta) \pm i\phi_{l}(\theta))$ expressed in polar form  for easier exponentiation, and thus:

\begin{widetext}
  \begin{align}
    1-\epsilon_{X, d_x \times d_z}(\theta) & =\exp(d_x \lambda_{d_z}(\theta))\cos(d_x \phi_{d_z}(\theta)) \approx (1-d_x\epsilon_{d_z}(\theta))(1-\frac{d_x^2\delta_{d_z}(\theta)^2}{2}) \\
    \delta_{X, d_x \times d_z}(\theta)     & = \exp(d_x \lambda_{d_z}(\theta))\sin(d_x \phi_{d_z}(\theta)) \approx (1-d_x\epsilon_{d_z}(\theta))(d_x \delta_{d_z}(\theta)),
  \end{align}

\end{widetext}
\noindent where the approximations are based on first-order expansions in $\epsilon_{d_z}, \lambda_{d_z}$ around zero, and thus they apply when these values are sufficiently small. In this low-noise regime, by further ignoring negligible third and second-order terms, we find that $\kappa_{X, d_x \times d_z}(\theta) \approx d_x \kappa_{d_z}(\theta)$, which means that the $X$-Shor code has the same coherence threshold as the phase-flip code. Thus, for $\theta$ values below $\theta_{th} = \pi/2$, the $X$-Shor code suppresses infidelity and coherence of the logical channel exponentially in $d_z$. This suppression is tempered by a factor of $d_x$, which, in the case of square codes $l=d_x=d_z$, is a linear slowdown.

In the case of the $X$-Shor code, our calculations reflect using a combined mininum weight recovery for each row of repetition codes, thus the threshold above is under MWPM decoding (see \cref{app:verifying formulae} for numerical evidence).

\begin{figure}[!htbp]
  \centering
  \captionsetup[subfigure]{justification=centering}
  \begin{subfigure}[b]{0.2\textwidth}
    \centering
    \includegraphics[width=\textwidth]{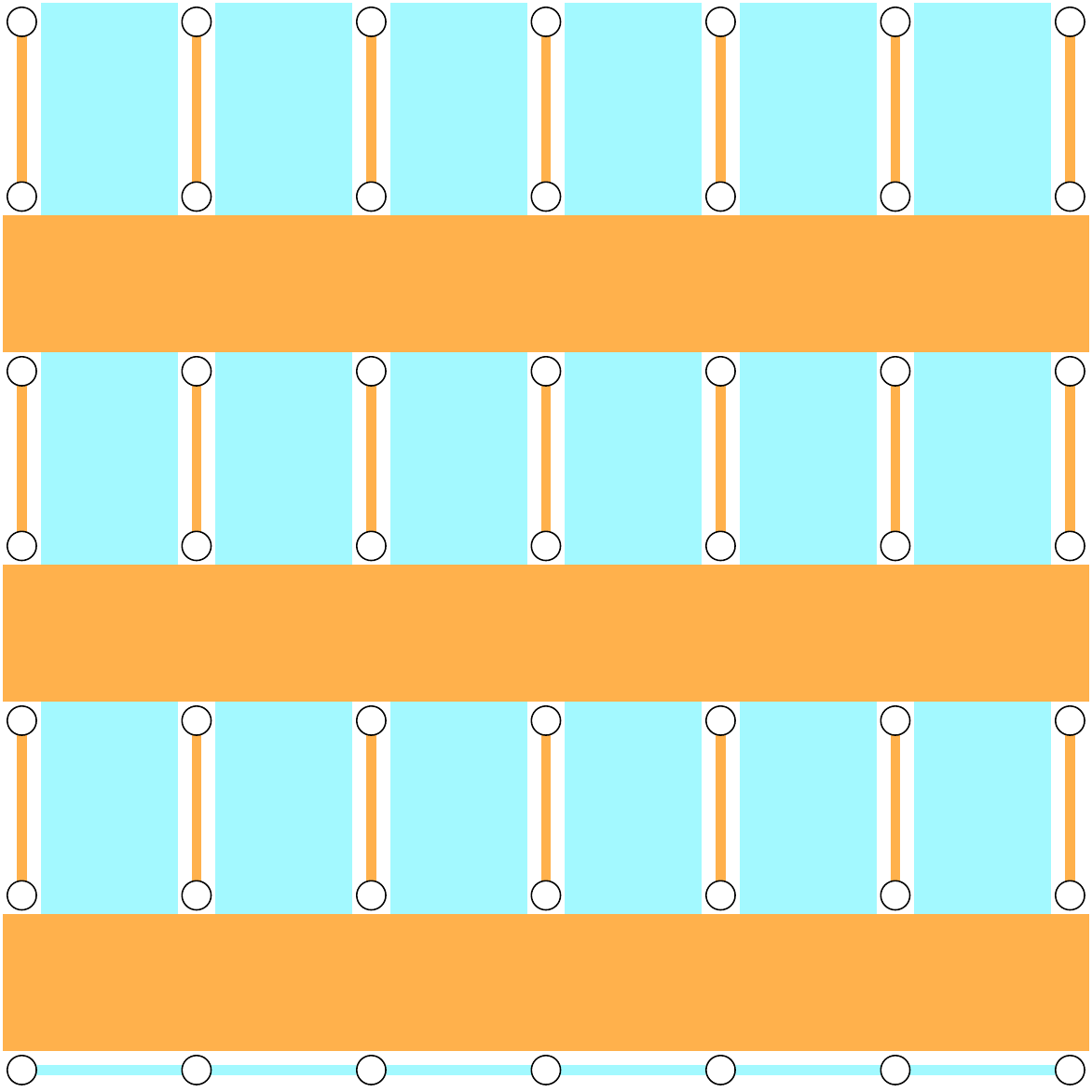}
    \label{fig:m2}
    \caption{}
  \end{subfigure}
  \hspace{10pt}
  \begin{subfigure}[b]{0.22\textwidth}
    \centering
    \includegraphics[width=\textwidth]{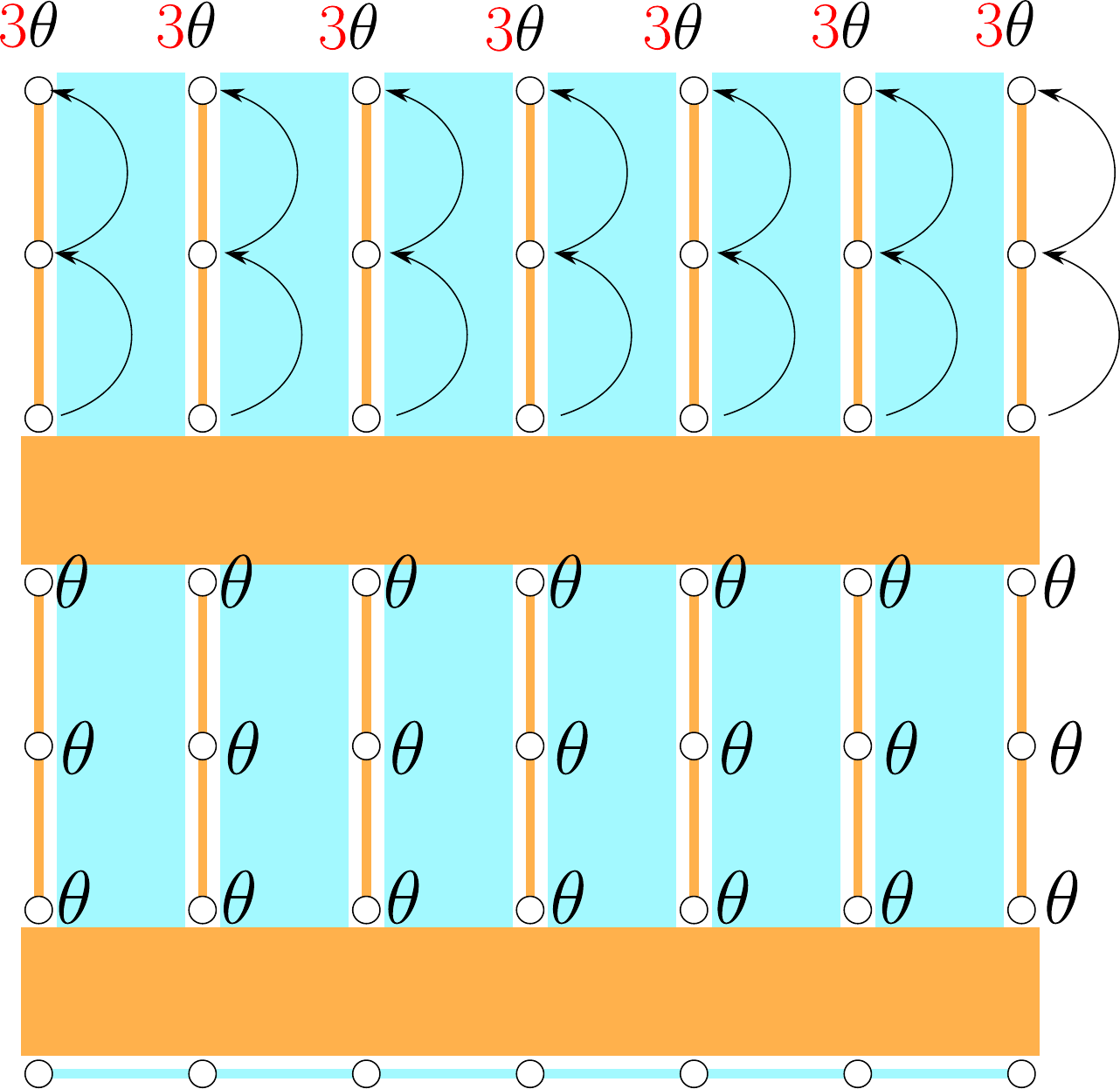}
    \label{fig:m3}
    \caption{}
  \end{subfigure}
  \caption{(color online) Two members of the $Z$-stacked Shor code family, $C_{7,2}$ (a) and $C_{7,3}$ (b). X stabilizers are blue (lighter) rectangles between two columns of qubits spanning from top to bottom, while Z stabilizers are orange (darker) rectangles between two rows of qubits spanning from left to right. In (b), the second $Z$-Shor block shows the uniform Z rotation by $\theta$. The top block shows the ``zipped'' view, which is equivalent to the uniform rotation up to the weight-2 $Z$-stabilizers. Here, qubits in the first row experience a $3\theta$ rotation while the other two rows do not participate.}
  \label{fig: repcompass}

\end{figure}

\subsection{Coherence threshold for the $Z$-stacked Shor codes} \label{ssec: $Z$-stacked shor code threshold}

We now define a generalization of the $X$-Shor codes, the family of \textit{$Z$-stacked Shor codes}, with members of length $l$ and block-height $h$, denoted $\mathcal{C}_{l,h}$, with $\mathcal{C}_{l,1}$ being the family of $l \times l$ square $X$-Shor codes. Consider the $l \times l$ square grid of qubits. In the first  $l_h \equiv \lfloor l/h \rfloor$ rows, we glue together $h \times l$ $Z$-Shor code blocks with weight-$2l$ $Z$ stabilizer generators. In the remaining $t_h \equiv l \mod h$ rows, we glue together $1 \times l$ $Z$-Shor code blocks (which are phase flip codes) with the $Z$ stabilizer generators. Two $l=7$ examples for $h=2$ and $h=3$ are displayed in \cref{fig: repcompass}. As the structure of these codes is similar to the $X$-Shor code, the logical channel's PTM is the product of the $Z$-Shor code PTMs. From before, we know that for each $Z$-Shor code block, we can use the parameters of the phase-flip code, just replacing $\theta$ with $h\theta$. Thus, the logical channel parameters for $\mathcal{C}_{l,h}$ are:
\begin{align}
  1-\epsilon_{\mathcal{C}_{l,h}}(\theta)= & \exp(l_h \lambda_{l}(h \theta) +  t_h \lambda_{l}(\theta))\times \nonumber       \\  &\cos(l_h \phi_{l}(h \theta) + t_h \phi_{l}(\theta)) \label{eq: exact compass eps}  \\
  \approx                                 & \exp(\frac{l}{h} \lambda_{l}(h \theta))\cos(\frac{l}{h}\phi_{l}(h \theta))       \\
  \delta_{\mathcal{C}_{l,h}}(\theta)=     & \exp(l_h \lambda_{l}(h\theta) + t_h \lambda_{l}(\theta) ) \times \nonumber       \\
                                          & \sin(l_h \phi_{l}(h\theta) + t_h \phi_l(\theta)) \label{eq: exact compass delta} \\
  \approx                                 & \exp(\frac{l}{h} \lambda_{l}(h\theta))\sin(\frac{l}{h}\phi_{l}(h\theta)),
\end{align}
\noindent where the approximations are valid at large $l$, where $l_h \gg t_h$, thus $l_h \approx \frac{l}{h}$. In the regime where $\epsilon_{l}(h \theta), \lambda_{l}(h \theta)$ are both sufficiently small, we can apply similar approximations as above for the $X$-Shor code. The resulting measure of coherence is approximately $
  \kappa_{\mathcal{C}_{l,h}} \approx \frac{l}{h} \kappa_l(h \theta)$. This result tells us that the coherence threshold for $\mathcal{C}_{l,h}$ is $\frac{\pi}{2h}$ and that the exponential suppression of infidelity and coherence is slowed down by a linear factor of $l/h$.

In the case of the $Z$-stacked Shor codes, not surprisingly, our calculations reflect performance under MWPM decoding  (see \cref{app:verifying formulae} for numerical evidence).

\subsection{Bounding the rotated surface code}

Under MWPM decoding, the infidelity thresholds $\pi/4$ and $\pi/6$ of the two families, $\mathcal{C}_{l,2}$ and $\mathcal{C}_{l,3}$, respectively, serve as bounding cases of the $\pi/5$ infidelity threshold of the surface code under MWPM decoding, which we reproduced as shown in \cref{fig:app combined}.

\begin{figure}[tpb]
  \includegraphics[width=0.45\textwidth]{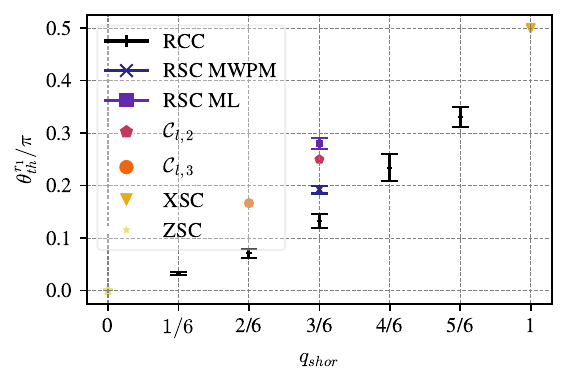}
  \caption{Infidelity thresholds $\theta_{th}^{r_1}$ of different compass codes. Random compass codes (RCC) are interpolated between the $Z$-Shor code (ZSC) and the $X$-Shor code (XSC) parametrized by the $X$-check density $q_{shor}$ using numerical experiments to estimate the average threshold. We display the analytically determined thresholds for the $Z$-stacked Shor code families $\mathcal{C}_{l,2}$ and  $\mathcal{C}_{l,3}$, and the numerical values for the rotated surface code (RSC MWPM) based on our simulation, which agrees with Bravyi et al.'s result \cite{bravyi_correcting_2018}. Finally, the ML decoding upper bound based on \cite{Venn_Behrends_Beri_2023} is displayed as RSC ML.}
  \label{fig:qshor}
\end{figure}

Finding a smoother transition might provide more insight into an analytical pathway toward a constant lower bound for the surface code threshold. We find that -- at least on average -- it is possible to smoothly interpolate between the extreme thresholds of the rotated surface code and the $Z$-Shor code thresholds using random compass codes of different $X$-check densities (\cref{fig:qshor}). We simulate random compass code families parameterized with $q_{shor}$, where $q_{shor}=1.0$ corresponds to the $X$-Shor code and $q_{shor}=0.0$ to the $Z$-Shor code. The parameter represents the portion of the coloring that comprises $X$-type stabilizers. In this parameter manifold $\mathcal{C}_{l,3}$ is $q_{shor}=1/3$ and both $\mathcal{C}_{l,2}$ and the rotated surface code are at $q_{shor}=1/2$. We evaluated the infidelity threshold for these codes numerically. Each data point uses 200 random codes and 200 samples for each distance across $d=9,13,17,21$. The threshold ranges are derived from plots in \cref{app:supporting data}. For these experiments, we implemented and generalized for compass codes a Majorana mode simulator invented by Bravyi et al. \cite{bravyi_correcting_2018}. Our simulator, \textit{msim}, is open sourced \cite{pato_msim_2023}. For MWPM decoding, we used PyMatching\cite{higgott2023sparse}.

\!\\

\section{Conclusions and outlook} \label{sec: conclusions}

2D compass codes are promising for exploring the open problem of proving a constant analytical lower bound for the rotated surface code threshold under uniform coherent $Z$-rotation, the worst-case scenario for coherent noise processes. We showed how the $Z$-Shor code can suppress coherent error in the $\sin(\theta) < 1/L$ regime, similar to the toric code \cite{iverson_coherence_2020}. Given that the $Z$-Shor code does not have a threshold and numerical evidence by Bravyi et al. \cite{bravyi_correcting_2018} points to a constant lower bound for the rotated surface code, there is room for improvement analytically as well.

We introduced the $Z$-stacked Shor code families that afford analytical analysis of the space. Two of these families yield thresholds under minimum weight decoding that provide upper and lower bounds to the rotated surface code's infidelity threshold. While an exact explanation is left for future work, the number of X and Z stabilizer generators being the same for the surface code and the $\mathcal{C}_{l,2}$ is suggestive of an underlying structural reason for the proximity of the two codes' thresholds.

Numerical investigations via the Majorana mode simulator can now be extended to compass codes with our open-source tool, {\it msim} \cite{pato_msim_2023}. These simulations show that it is possible to find structural properties that move the threshold on average. Still, the simple $X$-check density, $q_{shor}$, in itself, on average fails to have a high enough resolution of the threshold to fully explain the relationship between the two bounding cases of stacked $Z$-Shor codes and the rotated surface code.

Based on the results of Venn, Behrends, and Béri \cite{Venn_Behrends_Beri_2023} and the author's other work \cite{Pato_Miao_etal_2024} under ML decoding, the surface code is able to reach the threshold of $(0.28 \pm 0.01) \pi$. This is slightly above the threshold of $\mathcal{C}_{l,2}$ under MWPM decoding. We conjecture that an ML decoder will not be able to increase a higher threshold in the case of these codes, despite potentially achieving lower error rates. Thus, bounding the surface code under ML decoding using compass codes is left for future work.

Future investigations might identify other structural factors that, when changed, move the thresholds of compass code families in a strictly monotonic way. Uncovering such structural property or a combination of properties would help establish a constant analytical lower bound.

The authors thank D. Debroy and R. Calderbak for useful discussions. This work was supported by ARO/LPS QCISS program (W911NF-21-1-0005) and the NSF QLCI for Robust Quantum Simulation(OMA-2120757).
Code and data are available for download at \cite{pato_2024_11197690}.

\bibliographystyle{apsrev}
\bibliography{refs.bib}

\appendix

\newpage
\section{Supporting data for threshold calculations} \label{app:supporting data}

\begin{figure*}[htbp]
  \centering
  \includegraphics[width=0.8\textwidth ]{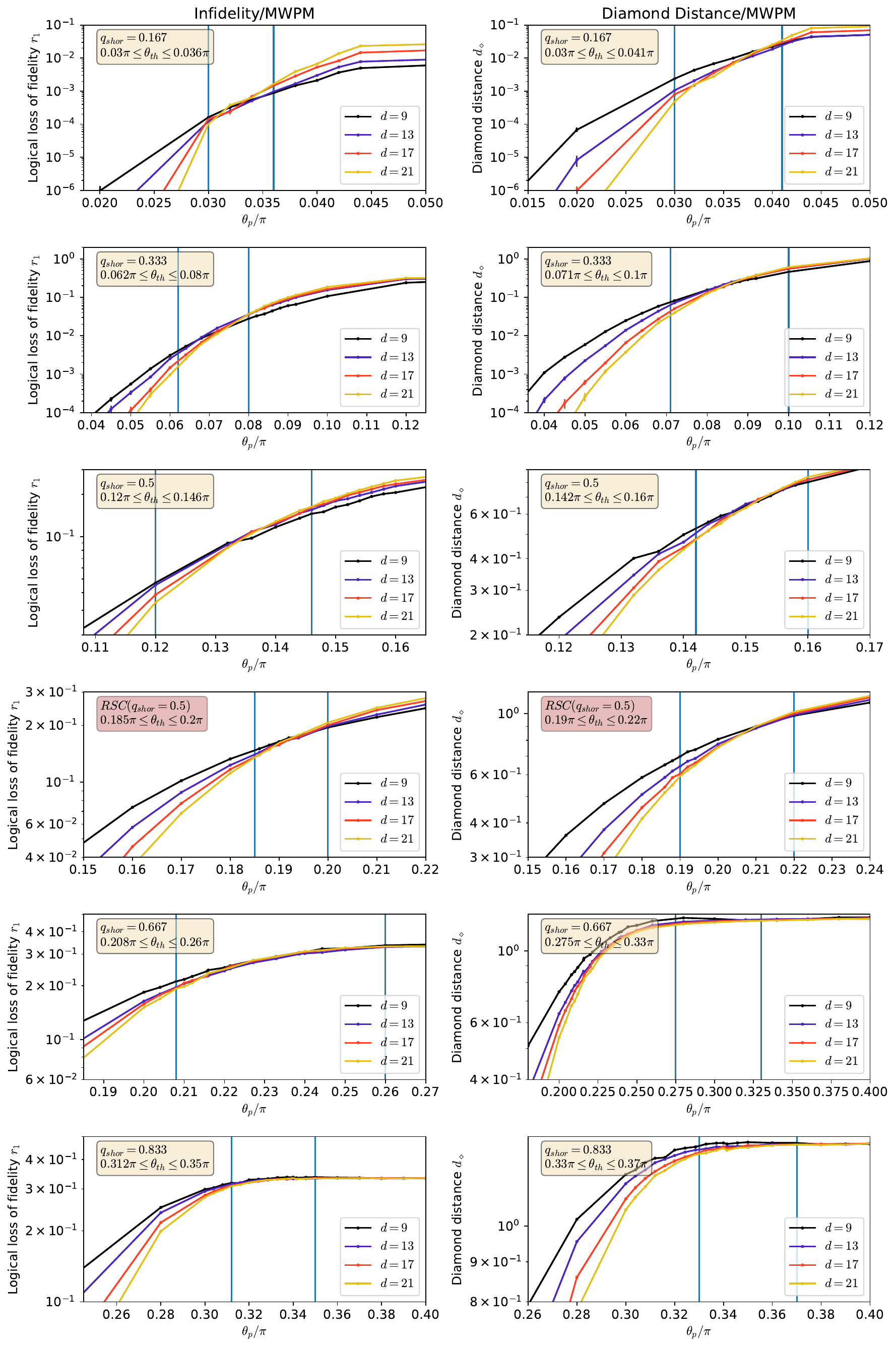}
  \caption{The underlying data for the threshold plot in the main text. We manually determine the upper and lower bounds for the thresholds by picking a regime that is clearly below threshold and one that is clearly above based on intersections observed at distances 9,13,17,21. The bounds are {displayed using} two blue vertical lines as well as the numerical values in the highlighted box. Rows are for the same code families, while columns represent the single-use channel fidelity $r_1$ (left column) and the diamond distance $d_\diamond$ (right column).}
  \label{fig:app combined}
\end{figure*}

In \cref{fig:app combined}, we display the data behind our threshold calculations based on random code families and the rotated surface code. For the random code families, each datapoint consists of 200 random families with the same $X$ check density and 200 samples from each families. Given the logical infidelity $1-\cos(\theta_s)$ for each measured syndrome with logical rotation $\theta_s$, we display the mean and the standard error for the observed $4\times10^4$ logical infidelities. We compare the average logical infidelity with the average logical diamond distance $2|\sin(\theta_s)|$ and standard error of the diamond distance. We observe slightly higher threshold values for the diamond distance.

\newpage
\section{Numerical verification of analytical results} \label{app:verifying formulae}

Here, we compare the predictions from our analytical results for repetition code based compass codes with numerical simulation data based on msim \cite{pato_msim_2023} using Pymatching \cite{higgott2023sparse} for MWPM decoding. There is perfect agreement between the two methods when the sample size is sufficiently large, verifying our calculations as well as our generalization of the Majorana simulator to the 2D compass codes.

\begin{figure*}[htbp!]
  \centering
  \includegraphics[width=0.8\linewidth]{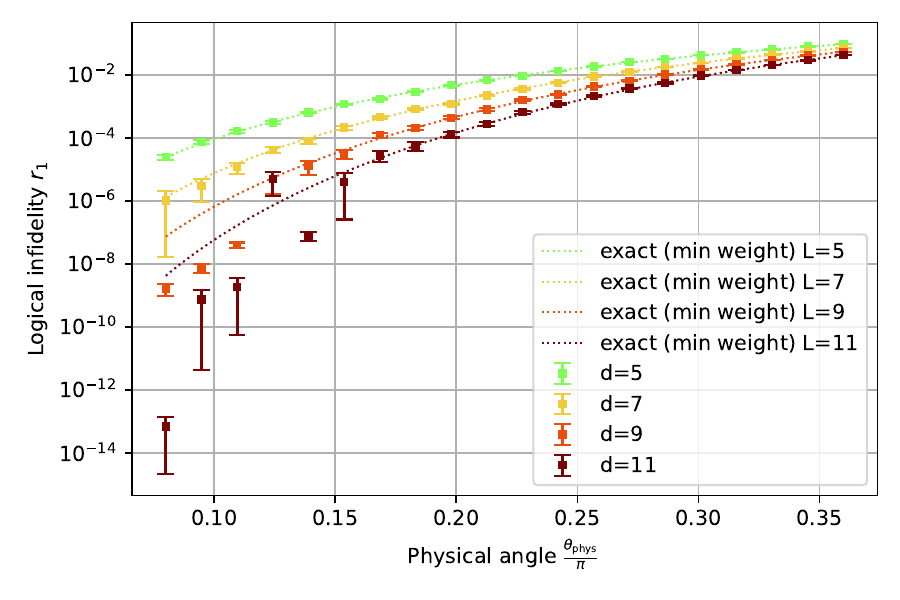}
  \caption{Logical infidelity $r_1$ for repetition codes. Exact (min weight) line shows the calculation using the repetition code based formulas in the text, without any approximations. For the numerical data, each data point is based on $10^4$ samples.}
  \label{fig:rep code infidelities}
\end{figure*}

\begin{figure*}
  \centering
  \includegraphics[width=.8\linewidth]{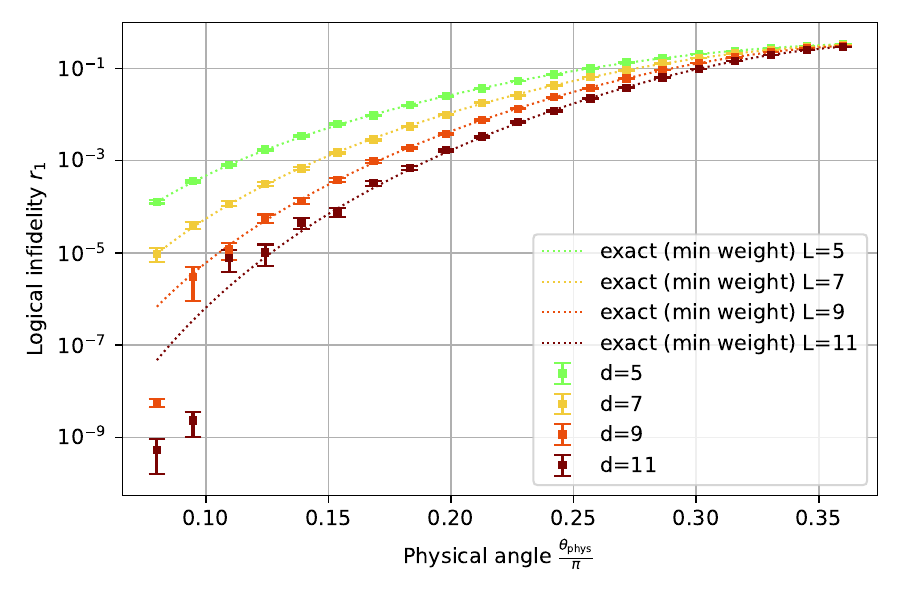}
  \caption{Logical infidelity $r_1$ for $X$-Shor codes. Exact (min weight) line shows the calculation using the repetition code based formulas in the text, without any approximations. For the numerical data, each data point is based on $10^4$ samples.}
  \label{fig:rep code infidelities}
\end{figure*}

\begin{figure*}
  \centering
  \includegraphics[width=.8\linewidth]{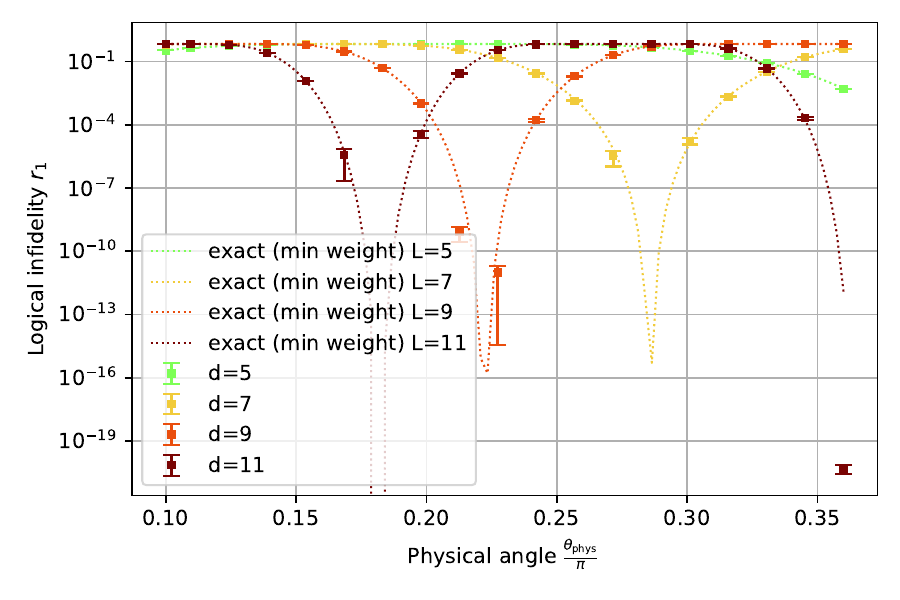}
  \caption{Logical infidelity $r_1$ for $Z$-Shor codes in a high error regime, in an oscillatory regime. Exact (min weight) line shows the calculation using the repetition code based formulas in the text, without any approximations. For the numerical data, each data point is based on $10^4$ samples.
    \label{fig:high error ZShor infidelities}}
\end{figure*}

\begin{figure*}
  \centering
  \includegraphics[width=.8\linewidth]{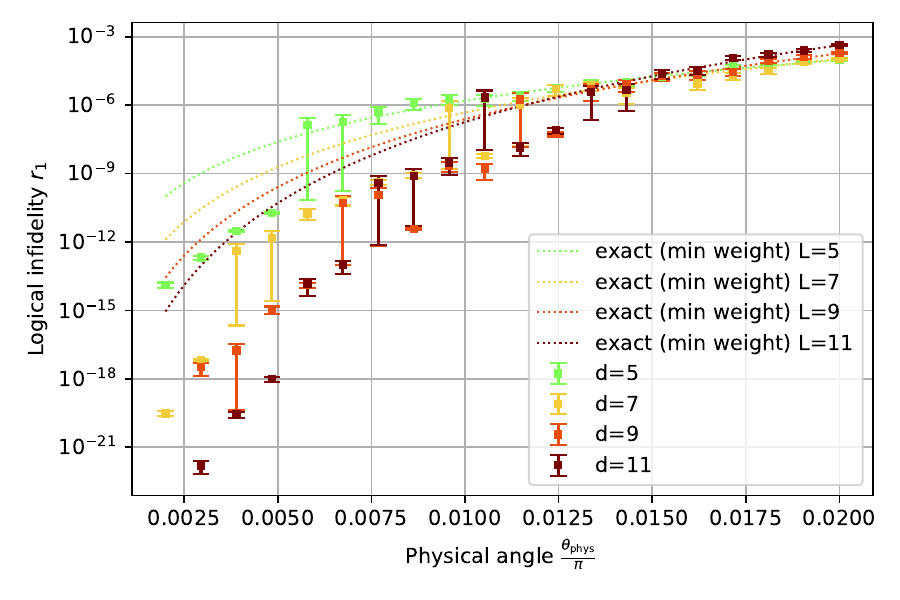}
  \caption{Logical infidelity $r_1$ for $Z$-Shor codes in a low error regime. It is clear that there is no true threshold, with the crossing between $d,d+2$ curves decreasing with increasing distance. Exact (min weight) line shows the calculation using the repetition code based formulas in the text, without any approximations. For the numerical data, each data point is based on $10^4$ samples. The low amount of samples are responsible for the large deviations at low error rates from the analytical formulae.}
  \label{fig:low error ZShor infidelities}
\end{figure*}

\begin{figure*}
  \centering
  \includegraphics[width=.8\linewidth]{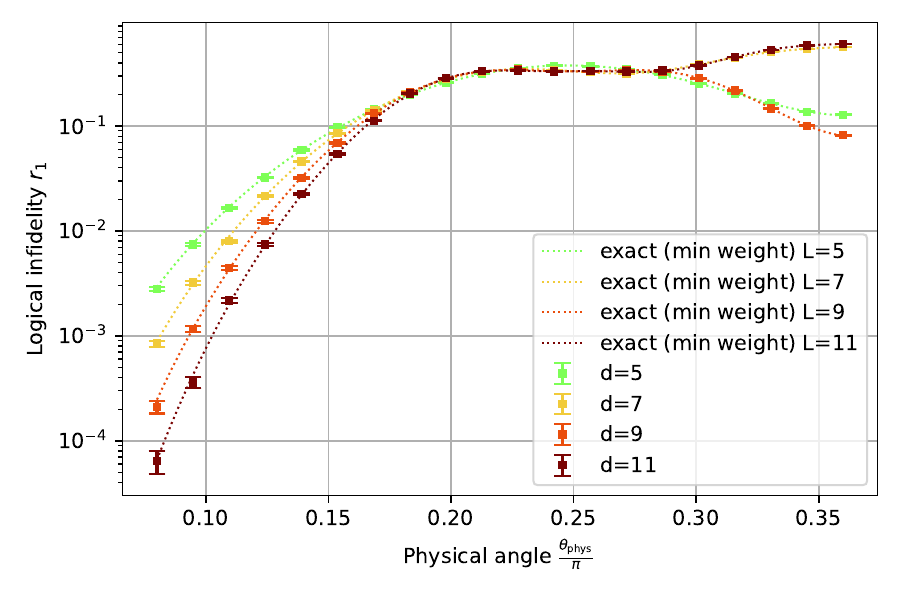}
  \caption{Logical infidelity $r_1$ for $Z$-Shor codes in a low error regime. It is clear that there is no true threshold, with the crossing between $d,d+2$ curves decreasing with increasing distance. Exact (min weight) line shows the calculation using the repetition code based formulas in the text, without any approximations. For the numerical data, each data point is based on $10^4$ samples. The low amount of samples are responsible for the large deviations at low error rates from the analytical formulae.}
  \label{fig:low error ZShor infidelities}
\end{figure*}

\begin{figure*}
  \centering
  \includegraphics[width=.8\linewidth]{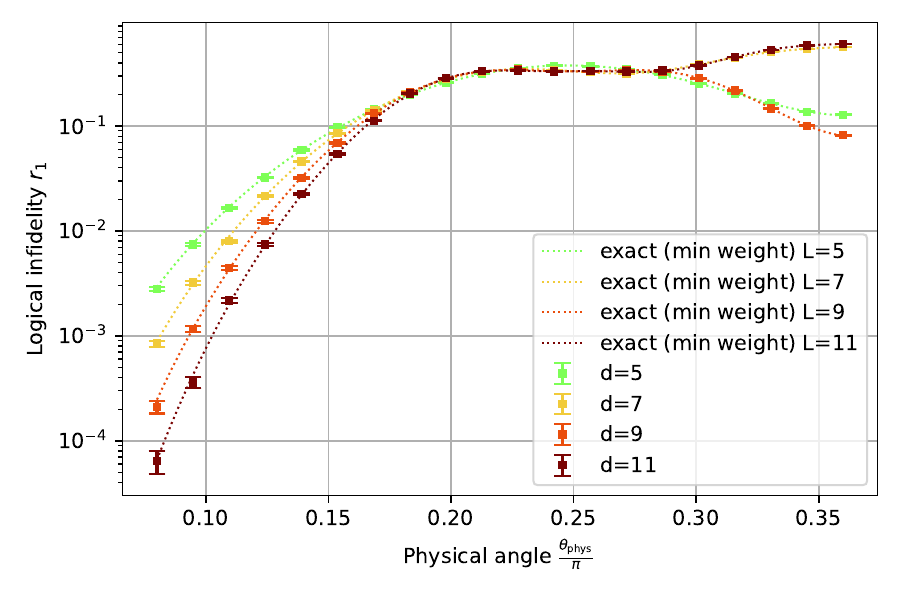}
  \caption{Logical infidelity $r_1$ for $C_{l,2}$ Z-stacked Shor codes. Exact (min weight) line shows the calculation using the repetition code based formulas in the text, without any approximations. For the numerical data, each data point is based on $10^4$ samples. The low amount of samples are responsible for the large deviations at low error rates from the analytical formulae.}
  \label{fig:low error ZShor infidelities}
\end{figure*}

\begin{figure*}
  \centering
  \includegraphics[width=.8\linewidth]{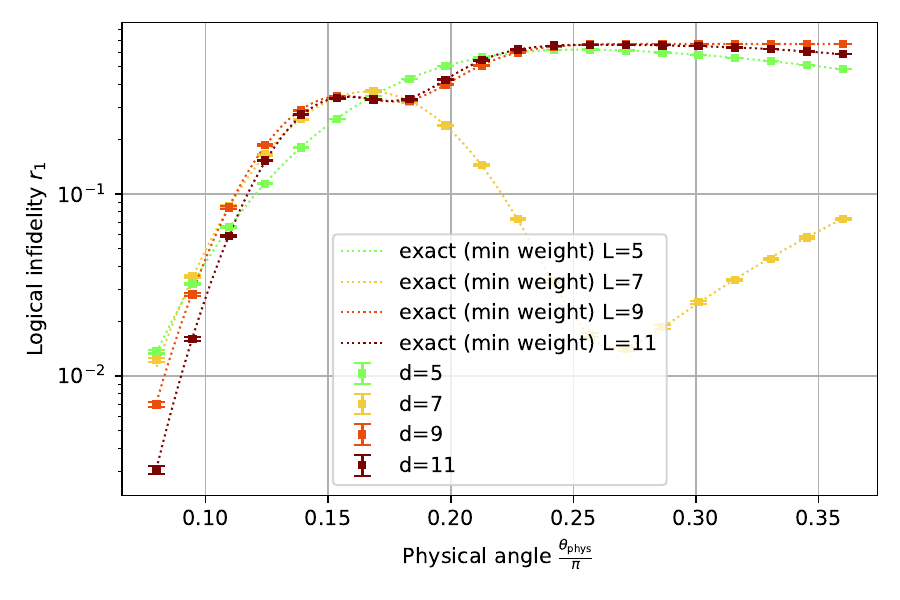}
  \caption{Logical infidelity $r_1$ for $C_{l,3}$ Z-stacked Shor codes. Exact (min weight) line shows the calculation using the repetition code based formulas in the text, without any approximations. For the numerical data, each data point is based on $10^4$ samples. The low amount of samples are responsible for the large deviations at low error rates from the analytical formulae.}
  \label{fig:low error ZShor infidelities}
\end{figure*}

\color{black}

\end{document}